\documentstyle[12pt]{article}
\begin{document}
\pagestyle{empty}
\begin{center}
{\Large\bf A multiwavelength study of the galaxy content and environmental  effects
in the dense group of galaxies around IC 65  } \\  
\bigskip
{\Large J. Vennik$^1$,  U. Hopp$^2$}\\
\smallskip
\begin{normalsize}
{ $^1$ Tartu Observatory, T\~oravere, Estonia}\\
$^2$~Universit\"atssternwarte M\"unchen, Scheiner Str. 1, D 81679 M\"unchen, Germany \\
\end{normalsize}
\end{center}                                                     

\section{Abstract}

In the framework of a larger program for studying the photometric properties of nearby groups
 of galaxies we have carried out surface photometry of four bright known members
 and a number of new possible LSB companions of the IC 65 (LGG 16) group of galaxies.
 The purpose of the present study is twofold: (a) a search for new LSB
 dwarf members and measuring of their global photometric characteristics, and
 (b) a search for possible effects of mutual interactions on the morphology and
 photometric characteristics of luminous and faint group members.
 
 This study is based on our $BRI$ CCD observations performed with the 1.23 m and 2.2 m
 telescopes at Calar Alto, which were supplemented by the deep $gri$ frames
 obtained from the Second Palomar Digital Sky Survey (DSS 2), and
 the NIR $JHK$ frames from the 2MASS database.
 In addition, we have used the HI imaging data from literature
 and the results of new pilot HI observations of selected dwarf-galaxy candidates
 made by W. Huchtmeier with the Effelsberg 100 m radiotelescope.

 In total, 105 galaxies have been detected with SExtractor software on the linearized DSS 2
 $60' \times 60'$ blue and red frames centered on LGG 16,
 and classified by their surface brightness, color and morphology. 
 We have discovered four new probable LSB dwarf members of the group,
 two of which have  been marginally detected in HI.

 The isophotal analysis in optical wavelengths revealed the presence of minor
 disturbances in the outer isophotes of some bright regular member galaxies.
 The HI emission appears generally more disturbed, compared to the emission in the
 optical passbands. The new dwarf companion candidates (with $-14.5 < M_B < -16.1$)
 generally contain blue star-forming regions
 with typical (Cousins) colours of $B-R = 0.65 \pm 0.15, R-I = 0.2 \pm 0.2$.
 The drift of optical isophotes together with the warping of HI isophotes
 outside the optical image of regular galaxies,
 as well as evidence for the enhanced star-forming activity in fragile
 dwarf galaxies indicate that this dense group of at least 8 late type
 galaxies is gravitationally bound and mutually interacting.

\begin{figure}[htp]
\unitlength0.1cm
\begin{picture}(150,200)
\includegraphics{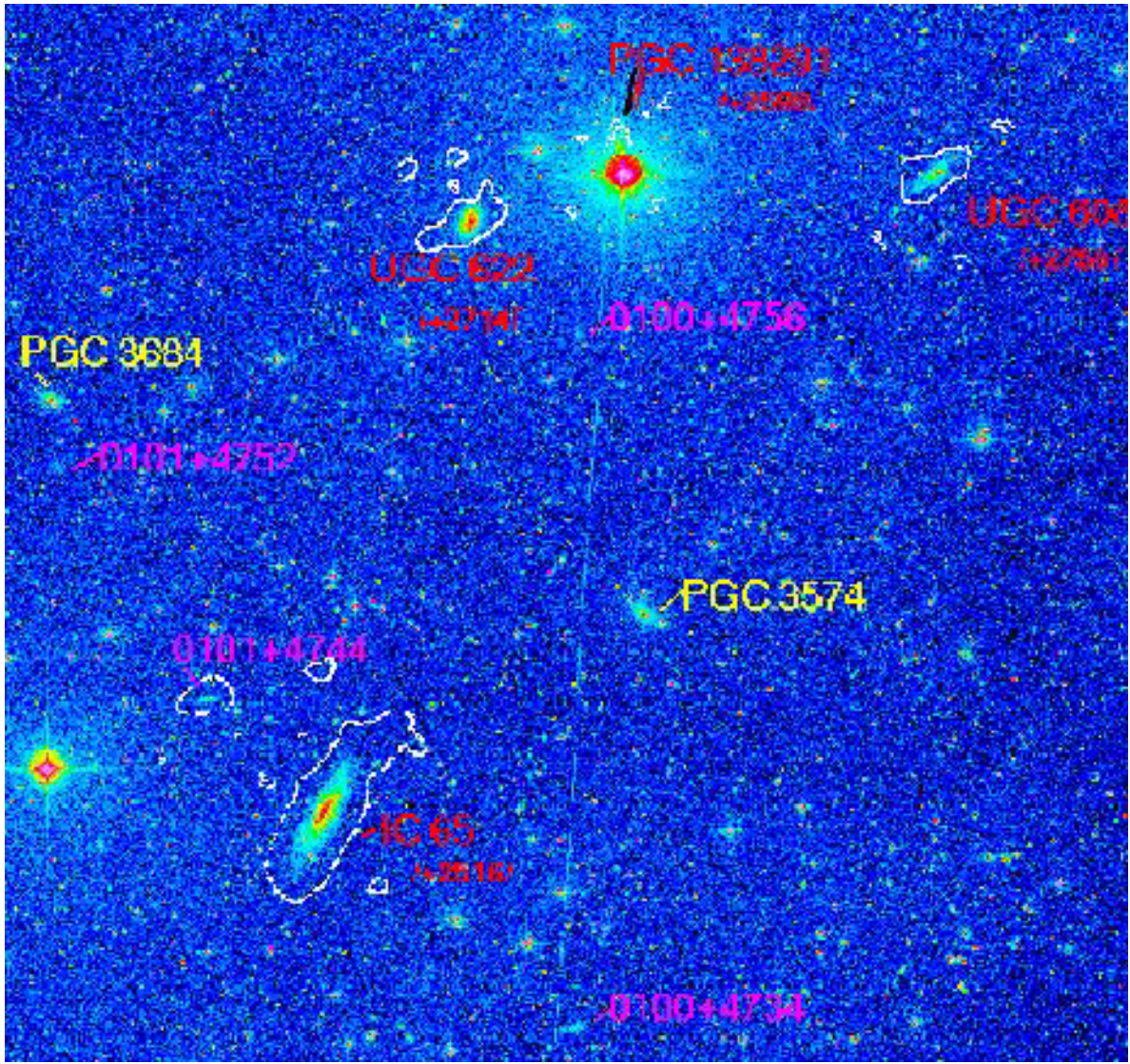}

\end{picture}
\caption{ DSS2 (POSS II) blue image of the IC 65 group of galaxies: four certain group members with
heliocentric redshits in brackets (red labels); 
four new probable LSB dwarf members (pink labels). The HI contours
(white) have been obtained from van Moorsel (1983) and correspond to the 2.5 - 3 $\sigma$ noise level (2.3 -
8.6 $\times~10^{20}$ atoms~cm$^{-2}$). Two early type galaxies (with yellow PGC numbers) are
probably located in the background of the group.
The size of the image is $55' \times~55'$. The north is at the top and the east is to the left.
}
\end{figure}

\section{Introduction} 

The IC 65 group of four late type galaxies (IC 65, UGC 608, UGC 622, and an "edge-on" galaxy - 
PGC 138291) 
has been studied by van Moorsel (1983) in the 
21 cm HI line with the WSRT. He noted: "each of these four galaxies 
show asymmetric features 
in their HI distribution, especially UGC 622. ... The presence of HI asymmetries and the small 
dispersion in radial velocities makes it likely that all detected four galaxies are 
members of one group" at the distance of $D = 38.5$ Mpc ($H_0 = 75$ km/s/Mpc is assumed).
Because of its unfavourable location close to the Zone of avoidance ($b^{II} \simeq -15^0$) 
this group was not registered in earlier catalogs of nearby groups, and has been registered as 
the Lyon Group of Galaxies (LGG) 16 (Garcia 1993). 
It fulfills selection criteria of the Hickson compact groups. 
Van Moorsel (1983) detected one more HI-rich LSB anonymous (A) galaxy located $\sim 5.0'$ to the 
NE of IC 65
(A~0101+4744 in the Fig.~1), having systemic velocity, which is close to the group mean.
This encourages us to look for further dwarf companions in the area of this 
particular group. 

\section{A search for new dwarf companion candidates}

We carried out a systematic search for new LSB dwarf member candidates of this particular group 
by utilizing the digitized POSS II (DSS 2) blue and red frames, which were previously linearized and 
photometrically calibrated by us. The searched area was $60' \times~60'$, 
i.e. 0.67 $\times$ 0.67 Mpc 
at the distance of the group.
The detection of objects was performed by means of running the SExtractor software 
(Bertin \& Arnouts 1996) in double-image mode on the 
blue and red DSS 2 frames. Galaxies were selected in the SExtracted catalog by several steps using 
the stellarity index, diameters and surface brightness (SB) characteristics. 
In a crowded environment of the IC 65 group the automatic 
selection was insufficient. We needed to take the morphologies into account. In effect, the final list 
of 105 galaxies was selected, which is believed to be complete for objects with diameters $\geq 2''$ 
at the 25.5 $B/\Box''$ isophote.

\begin{figure*}[h]
\unitlength0.1cm
\begin{picture}(160,70)
\includegraphics{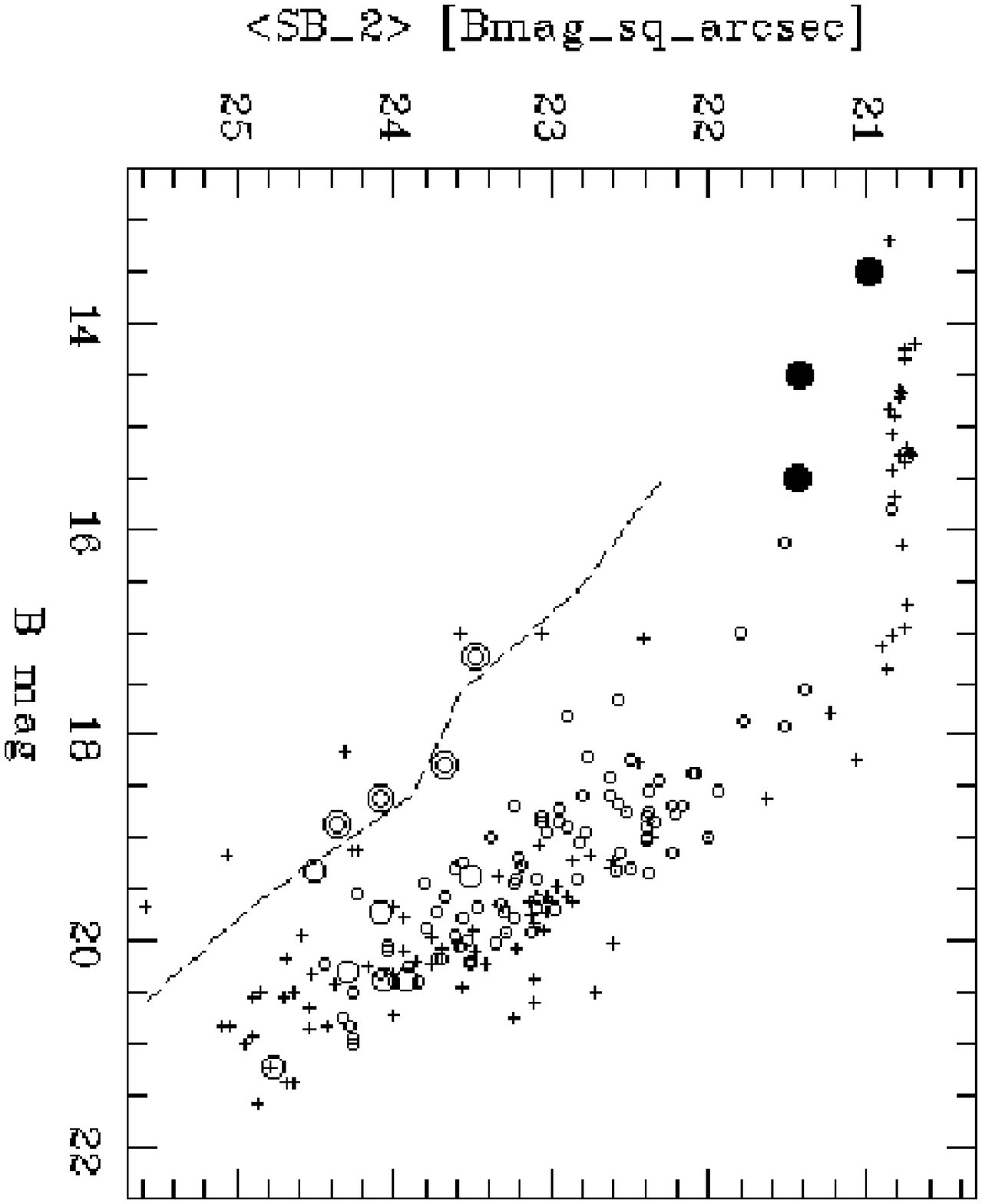}
\includegraphics{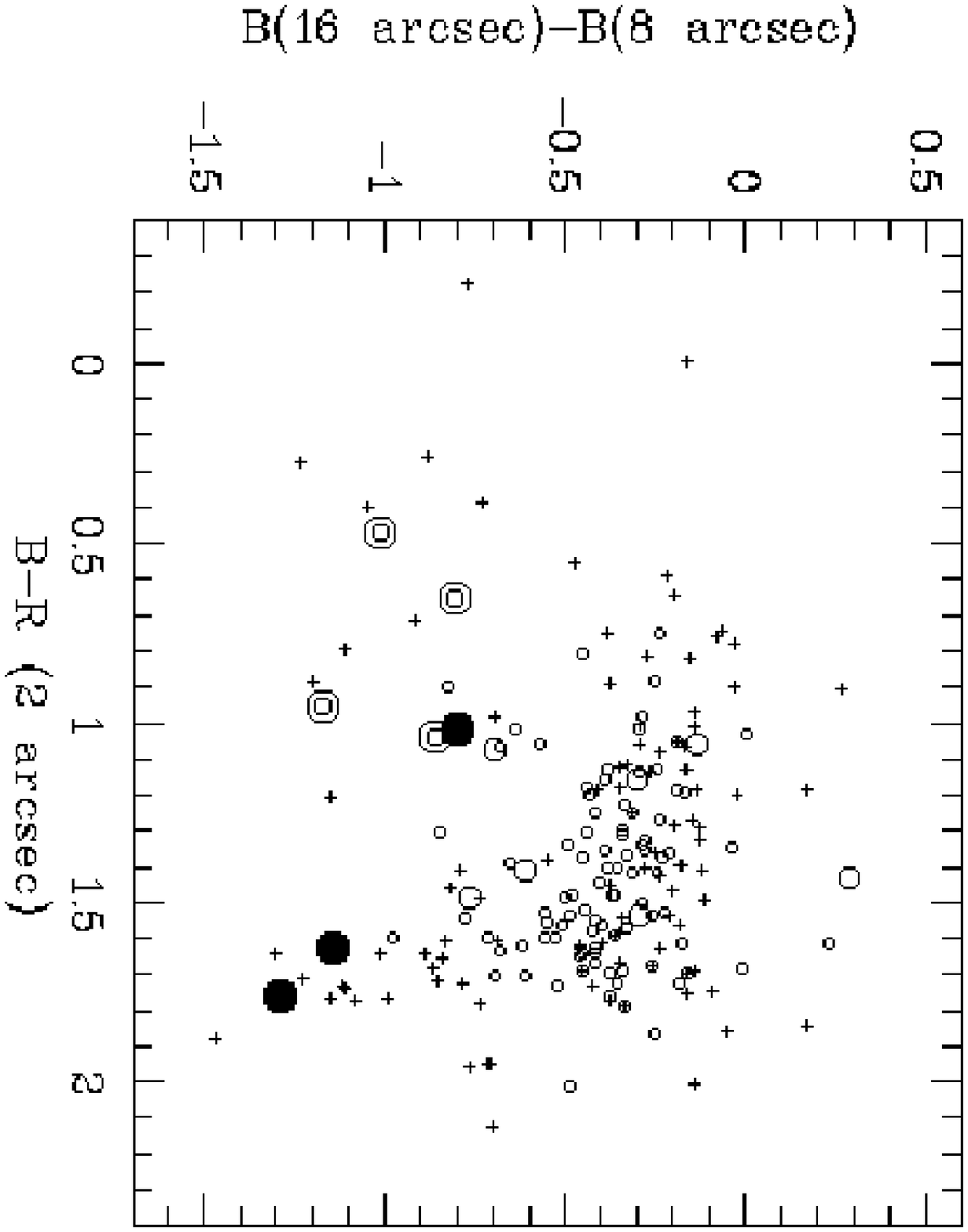}
\end{picture}
\caption{Classification of SExtracted objects in the field of the IC 65 group:
{\it left)} Mean surface brightness within $2''$ aperture $<SB\_2>$ as a function of
 total $B$-magnitude. The dashed line is the median predicted position of dwarf galaxies
, obtained from Ferguson \& Binggeli (1994, Fig.~3), and scaled for the IC 65 group.
{\it right)} Light concentration parameter $B(16'')-B(8'')$ versus central $B-R$
color index. Coding: {\it filled hexagons} - luminous certain group members; {\it double circles} -
large LSB galaxies -- probable new dwarf members of the group; {\it large circles} - other LSB galaxies
in the field; {\it small circles} - HSB galaxies in the field; {\it crosses} - bright stars with
$B < 15$ mag.
}
\end{figure*}

In order to disentangle the possible new dwarf members of the group from distant field galaxies, 
we applied two criteria: 1) the Binggeli's (1994) empirical relation between the central SB and absolute
 magnitude, common both for dE and dIrr galaxies (Fig.~2a), and 2) the empirical light concentration
 parameter, as proposed by Trentham et al. (2001), versus SExtracted colour index (Fig.~2b).   
As a result, we have selected four LSB irregular galaxies, which show the lowest light concentration 
(i.e. the largest scale lengths), the bluest color ($0.45 < B-R < 1.05$), and fit the empirical Binggeli's 
relation for dwarf galaxies. There are 7 more galaxies with LSB morphologies, but all of them 
with $B > 19.0$. At fainter magnitudes and/or smaller diameters we actually lose the ability to 
distinguish possible group members from the background galaxies on SB grounds. \\
Table~1 lists the basic data for certain group members and for new selected anonymous (A) 
probable dwarf members. All magnitudes are corrected for Galactic absorption.

\begin{table*}[h]
\caption{\normalsize Basic data of the studied galaxies}  
\begin{flushleft}
\begin{small}
\begin{tabular}{llllcrllll}
\hline
 Galaxy     & Type     & $D \times d$& $P.A.$ & $V_{hel}$ & $B_T$ & $M_B$ & $B-R$ & grd$^{1)}$ & $R-I$ \\
 (1)        &     (2)      &    (3)      &  (4)     &   (5)       &  (6)   &     (7)   &  (8)  &  (9) & (10) \\
\hline
\\
 UGC 608    & SABdm & 2.$^{'}$0~~0.$^{'}$9 & 128$^o$ & 2755 & 14.34 & -18.26 & 0.74 &0 & 0.3:  \\
 PGC 138291 & S(dm)    & 1.3:~~0.3: & 165  & 2598 & 16.5: & -16.4: & 1.1: & 0 & 0.2:  \\
 A0100+4756 & dIrr     & 0.6:~~0.4: & 30:  &      & 18.5: & -14.4: & 0.75 & 0 & 0.2: \\
 A0100+4734 & dIrr     & 0.8:~~0.3: & 108: &      & 16.86 & -16.1: & 0.74 & + & 0.25 \\
 UGC 622    & Scd:     & 1.1~~~0.7  & 159  & 2714 & 13.76 & -19.21 & 1.27 & +: & 0.61  \\
 IC 65      & SAB(s)bc & 4.4~~~1.2  & 153  & 2614 & 12.69 & -20.24 & 1.07 & 0 & 0.56  \\
 A0101+4744 & dIrr     & 0.7:~~0.2: & 100  & 2760 & 17.56 & -15.4 & 0.78 & + & 0.1: \\
 A0101+4752 & dIrr     & 0.4:~~0.3: & 20:  &      & 18.0: & -14.9: & 0.66 & + & 0.2:  \\
\hline
\end{tabular}
\end{small}
\end{flushleft}
\vspace{-3mm} 
$^{1)}$ $B-R$ radial color gradient (grd): positive gradient (+) and zero gradient (0)
\end{table*}

\section{Observations}

Broad-band $B, R$ and $I$ frames were taken during two observing periods in 
1995 and 1999 with the Calar Alto 1.23 m telescope and 1024 $\times$ 1024 pixel CCD camera, yielding 
a field of view of 8.7$' \times 8.7'$ with a scale of 0.51$''$/px. The exposure times were 
100 - 600 sec with seeing conditions typically between 1.5$'' - 2.5''$ (FWHM). 
For magnitude calibration and color correction, standard stars were
observed simultaneously in the star cluster NGC 7790 (Christian et al. 1986). 
Deep images (1800 sec in $B$ and 1200 sec in $R$), which were taken in 1995 under nonphotometric 
conditions, were calibrated by means of comparing total fluxes of a sample of nonsaturated stellar 
images in the photometric and nonphotometric frames.

\section{Data reduction}

After standard correction procedures the CCD frames were smoothed by means
of the adaptive filtering technique described in Lorenz et al. (1993) and
run within the MIDAS environment. Next, the sky background was fitted by a tilted plane, using the 
least-squares method. After background subtraction the galaxies were interactively cleaned from 
disturbing objects using the interactive polygon editor.

The surface brightness (SB) distribution was analyzed by two
methods. First, we calculated the equivalent light profile by slicing
the smoothed galaxy image at predefined intensity levels and counting
the pixel intensities in the areas between successive isophotes. The
differential counts reduced to a unit area give the azimuthally
averaged intensity profile of the galaxy as a function of the equivalent
radius. This procedure permits to enhance  
signal-to-noise ratio for the faint outskirts of the galaxies and 
to proceed the SB profile until $\sim 26.5 B/ \Box''$. A set of
isophotal, effective and asymptotic photometric parameters was
determined on the basis of the equivalent light profiles and 
light growth curves. 

The second approach should receive some information about the
two-dimensional structure. For this purpose, we used the ellipse
fitting algorithm of Bender \& M\"ollenhoff (1986) as available in
MIDAS. We obtained a set of radial profiles - surface
brightness (SB), axis ratio (b/a), position angle (PA) and decentering ($x-x_0$ and $y-y_0$) - in each
particular passband. The color-index profile was calculated by means
of combining these particular SB profiles.
 
\section{SB profile description}

The SB profiles depend on the morphological type and bear some information about the possible
environmental influences on the dynamical evolution of galaxies. 
Preliminary inspection of the derived SB profiles
reveals that those of the LSB dwarf galaxies show minor deviations from the pure exponential mostly
in their inner parts, because of the occurrence of star-forming knots, if any.
The outer (generally rather
noisy) profile might represent the older, underlying disk of the dwarf galaxy and therefore
provides physically meaningful parameters when fitted by an exponential density law.\\
The luminous group members show evidence of multiple components in their SB profiles. However, the
conventional bulge/disk decomposition looks straightforward only for the parent
galaxy IC 65 and the late type LSB galaxy UGC 608, both of which show evidence of a faint bulge and/or
a bar component. Another bright member galaxy UGC 622 shows a typical type II profile (Freeman 1970), 
and the "edge-on" galaxy is evidently a pure late type disk galaxy.
The obvious first step is to fit the SB profiles with the exponential intensity law
$\mu (r) = \mu_o + 1.086~r/h$.
The scaling parameters ($\mu_o, h$) have been obtained from the SB profiles using the "marking the 
disk" method. The disk fitting range was defined for each galaxy individually in 
outer part of the profile. 
In the case of  IC 65 and UGC 608, both of which show reliable excess light above the exponential disk
model, this extra light was modeled with a Sersic (1968) power law 
$\mu (r) \propto (r/h)^{1/n}$, which has been found  proper when describing the light 
distribution of spiral bulges (e.g. Andredakis et al. 1995). 

\section{The main results of surface photometry}

The main photometric results are presented in the form of a) a set of 2D images (broad-band optical 
and NIR image, color-index image, HI contour image (when available); b) 1D radial profiles (SB and   
color-index profiles, etc.); c) tables with conventional photometric and exponential model parameters. 
The 2D images show important morphological aspects such as the bulge contribution, the presence of bars 
and the shape of spiral arms. We apply contours of the Laplacian-filtered image 
in order to reveal specific morphological features in luminous parts of the galaxies. 
For the bright galaxies we extracted individual $J, H$ and $K$ frames from the 2MASS database, and
showed the $JHK$ composite image, which characterizes the distribution of
old stellar populations in these galaxies. 
We also showed the color-index image of each bright galaxy and of those dwarf galaxies for which 
reliable colors could be obtained above the noise level. 
It is especially useful to visualize the stellar population and dust distribution. 
The parameters describing the axis-ratio ($b/a$) and the orientation ($P.A$) of the galaxy
isophotes were calculated as an average obtained from fitting ellipses typically to the 
range of isophotes 24 - 25.5 $B/\Box''$. In order to quantify the irregularities 
in optical morphology of the outer disk of regular galaxies, we calculated the {\it decentering degree}
 as the displacement of the center of the most external isophote with respect to the
luminosity center, normalized to the last measured radius (Marquez \& Moles 1999). 

Here we present an abridged version of the obtained results for the parent galaxy IC 65 (Fig.~3) as well 
as for all four new probable dwarf members of the group (Figs.~4 - 7). More results for all studied  
galaxies are given in Vennik \& Hopp (2004).

\section{Discussion and summary}

 To our present knowledge, groups of galaxies are a proper place to study interactions since the galaxy 
encounters in groups occur at lower velocities, compared to the clusters of galaxies, and may lead to 
efficient merging events. Dominant luminous galaxies in the group can potentially reveal the history 
and evolutionary status of the respective system. Old groups usually contain one or several luminous 
ellipticals, which may have formed in merging episodes and could have structures reminiscent of past 
interaction events. On the other hand, pure spiral groups should be dynamically younger aggregates. 
Their members could reveal ongoing interaction through morphological distortions and in their possibly 
triggered star-formation activity (Tanvuia et al. 2003). 
The dynamically youngest groups are good places to look for 
galaxies evolving at $z \sim$ 0. 

The analysis (of available photometric and kinematic data) for the IC 65 group of galaxies 
leads to the following results:

\noindent 
$\bullet$ This particular compact group of four luminous late type spiral galaxies is well isolated 
both in redshifts and in projection to the sky. 
We could identify four LSB dwarf companion candidates on deep DSS 2 frames according to their SB, 
color and morphology. \\
$\bullet$ Available HI imaging data show that all bright members and at least one new dwarf companion 
are rich in HI with HI-mass to blue luminosity ratios in the range of 0.4 -- 1.2.
Their HI halos are typically twice as extended as the optical Holmberg radii of individual galaxies.
The outer HI isophotes (with column densities in the range of 2.3 -- 8.6 $\times 10^{20}$ atoms~cm$^{-2}$) 
generally appear disturbed, especially in IC 65 and in UGC 622 (see Fig.~1).\\
$\bullet$ Optical morphology of bright galaxies generally appears to be regular, with minor 
disturbances in outer isophotes: e.g. in IC 65 the decentering degree is $\sim$ 8.2 \% and 
in UGC 622 -- $\sim$ 6.3\%, which are significantly larger than the average decentering of 2.4 $\pm$ 2.7 
\%, found by Marquez \& Moles (1999) for isolated spirals. \\
$\bullet$ All bright group members (except PGC 138291, which we could not study in such 
detail) consist of many blue star-forming knots and plumes, especially UGC 608.\\
$\bullet$ UGC 622 is a Scd galaxy without a classical bulge, but containing a tiny bright nucleus. 
It shows type II SB profile with two exponential sections of different scale lengths.\\
$\bullet$ New probable dwarf companions are all of irregular appearance with blue knots, 
especially A~0101+4744 and A~0100+4734, for which a chain of blue SF patches give a cometary 
appearance. \\
$\bullet$ Significant positive color gradients were found in LSB dwarf galaxies of short 
scale lengths, and they are related to starbursts. Late type dwarfs generally show a stellar 
population gradient, i.e. 
younger and bluer stars are more centrally concentrated and redder stars dominate the underlying older 
disk component. 

Morphologically disruptive interactions are expected to be more common in low-velocity-dispersion 
($\sigma_v$) less evolved 
groups such as the Local Group and the M 81 group (Zabludoff 2001). 
The IC 65 group of galaxies obviously 
belongs to the class of less evolved groups: with its $\sigma_v$ = 70 km/s, observed drift in optical 
isophotes and 
obvious tidal features in HI outer isophotes, as well as with the evidence of enhanced SF activity in 
fragile dwarf companion candidates let us conclude that the IC 65 group of galaxies is probably dynamically 
young and possibly at the stage of its collapse.

\section{ References} 

Andredakis, Y.C., Peletier, R.F., Balcells, M., 1995, MNRAS, 275, 874 \\ 
Bender R., M\"ollenhoff C., 1987, A\&A 177, 71\\ 
Bertin, E., Arnouts, S., 1996,  A\&AS, 117,393 \\
Binggeli B., 1994, in: G.Hensler, C.Theis, J.S.Gallagher (eds), "The
panchromatic view of galaxies", Editions Frontier, p. 173\\
Christian C.A., Adams M., Barnes J.V., Butcher H., Mould
J.R., Siegel M., 1985, PASP 97, 363\\ 
Ferguson, H.C., Binggeli, B., 1994, A\&ARv., 6, 67 \\
Freeman, K.C, 1970, ApJ, 160, 811 \\ 
Garcia, A.M., 1993, A\&AS, 100,47 \\ 
Lorenz H., Richter G.M., Capaccioli M., Longo G., 1993, A\&A 277, 321\\ 
Marquez, I., Moles, M., 1999, A\&A, 344, 421 \\ 
Sersic, J.L., 1968, Atlas de galaxias australes. Cordoba \\
Tanvuia, L., Kelm, B., Focardi, P., Rampazzo, R., Zeilinger, W.W., 2003, AJ, 126, 1245 \\ 
Trentham, N., Tully, R.B., Verheijen, M.A.W., 2001, MNRAS, 325, 385 \\ 
van Moorsel, G.A., 1983, A\&AS, 54, 1 \\
Vennik, J., Hopp, U., 2004, in preparation \\
Zabludoff, A., 2001, ASP Conf. Proc. 240, 547 \\  

\newpage 
\begin{figure*}[hp]
\unitlength0.1cm
\begin{picture}(155,130)
\includegraphics{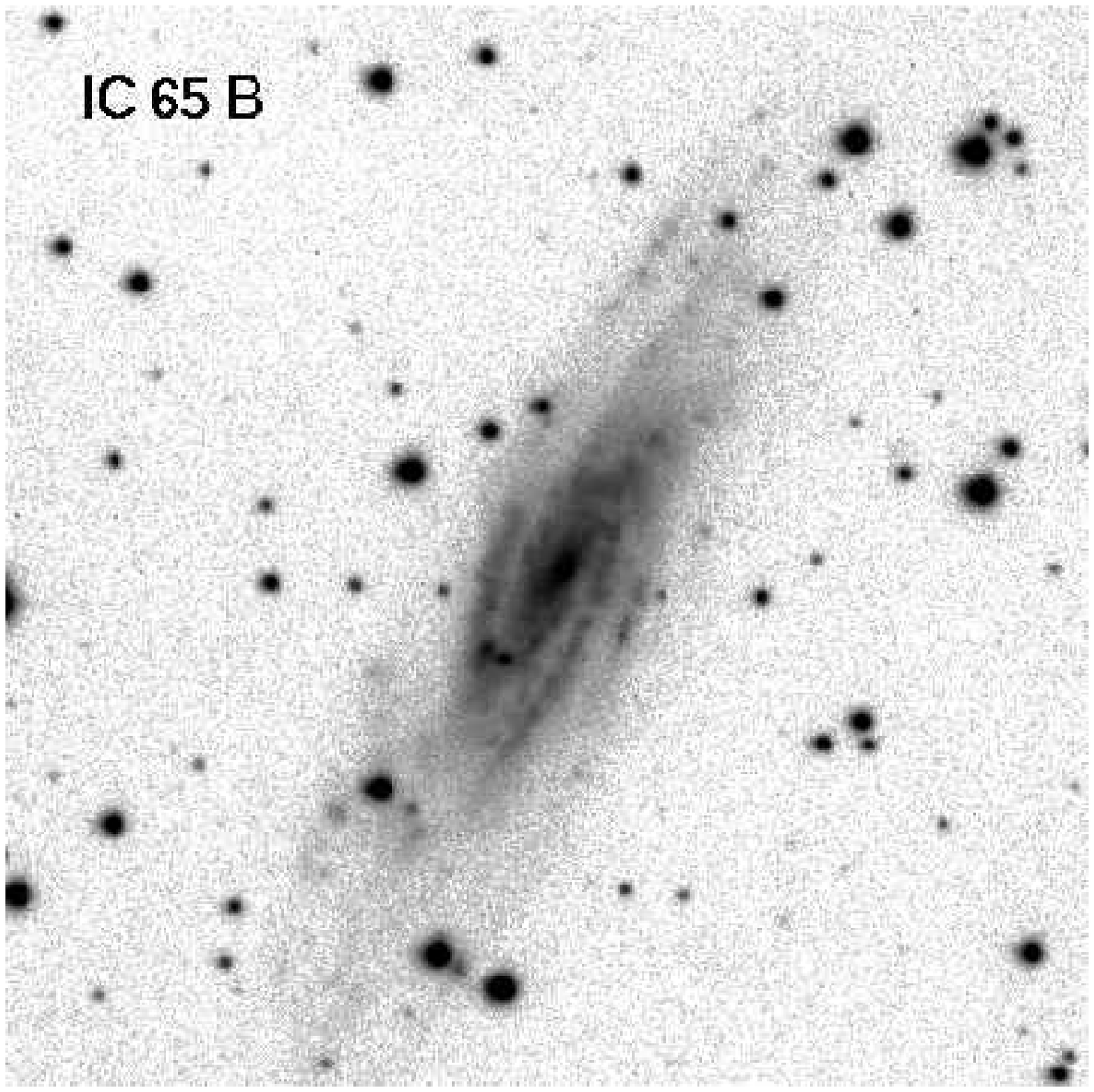}
\includegraphics{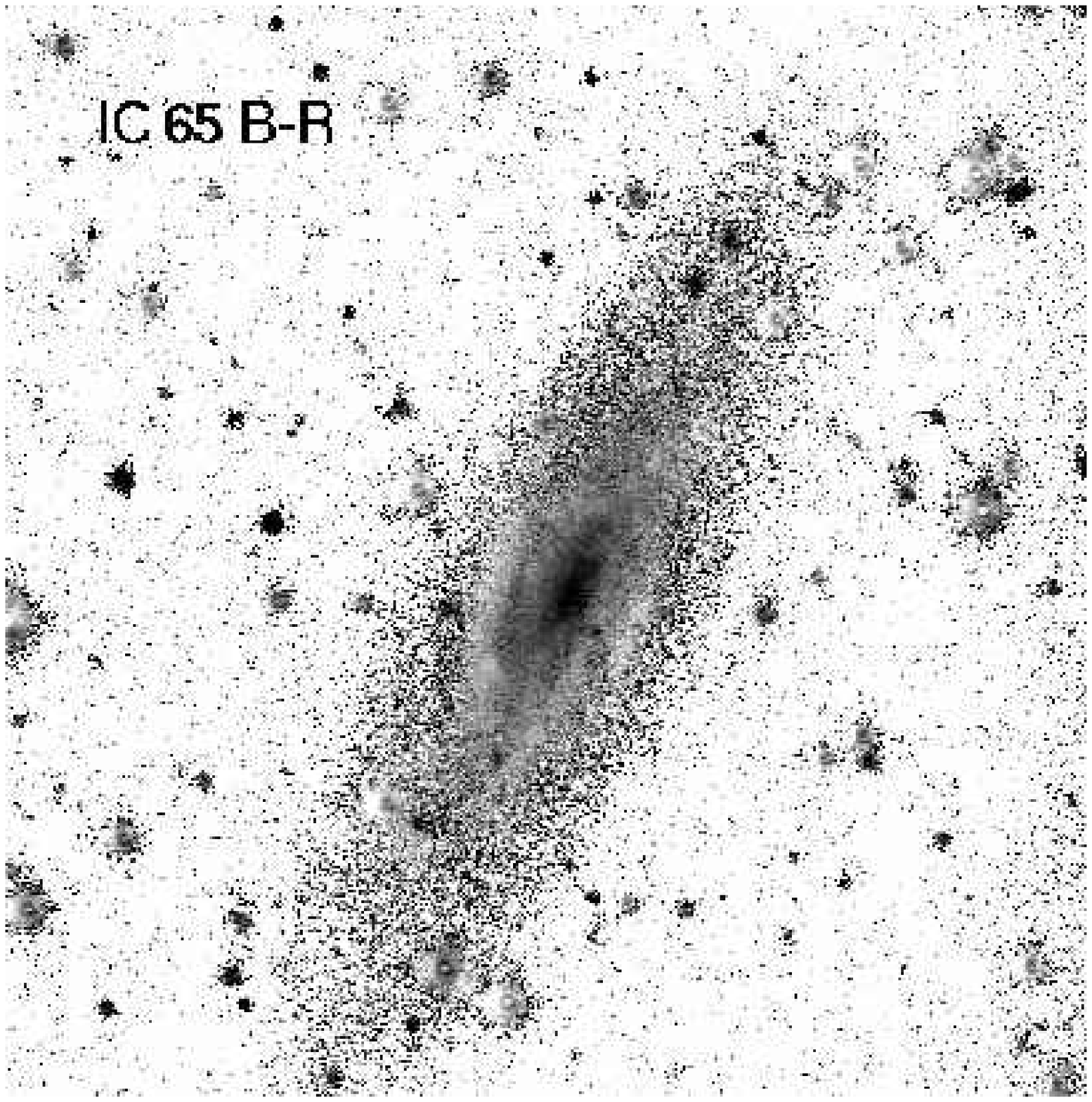}
\includegraphics{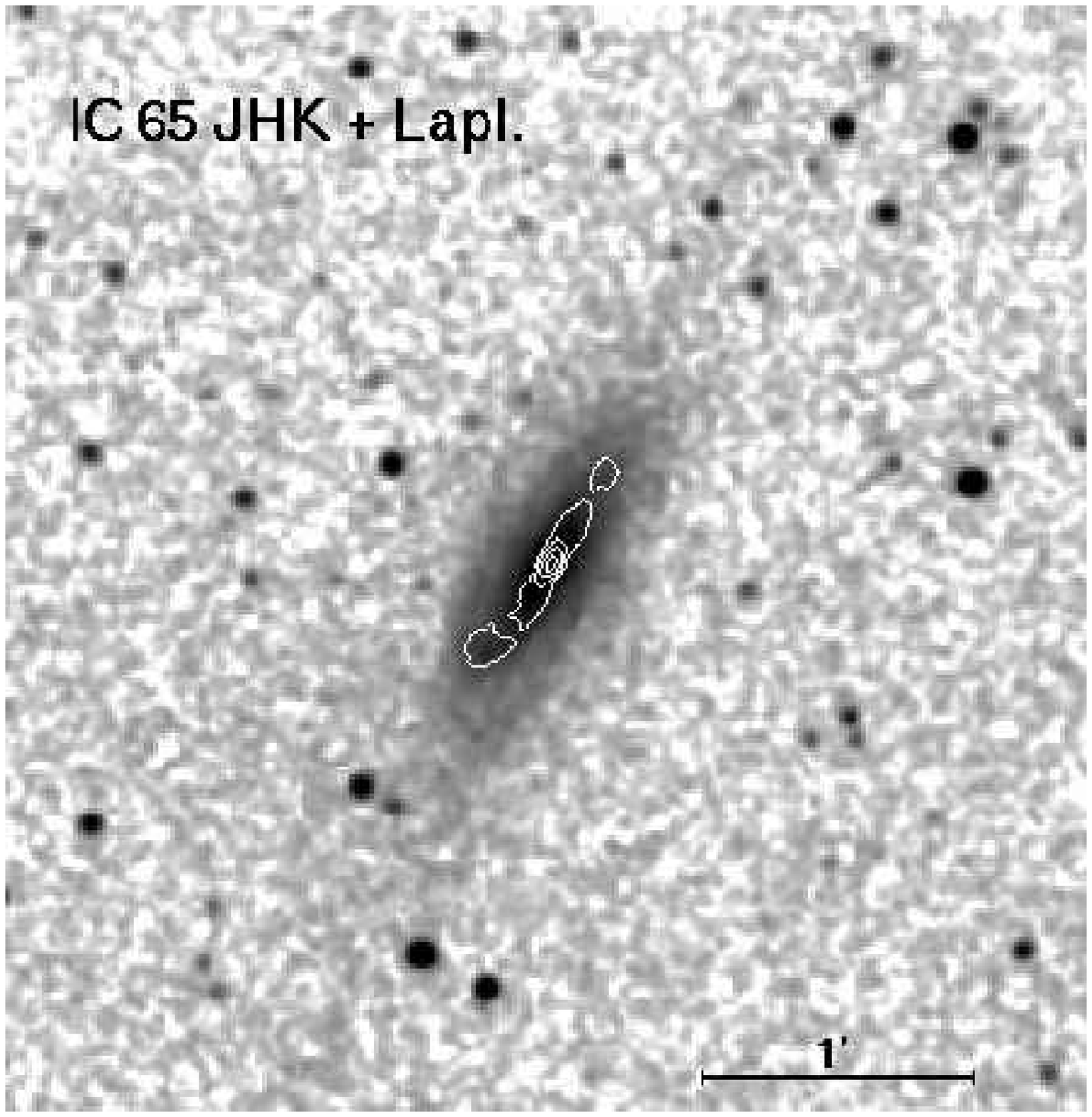}
\includegraphics{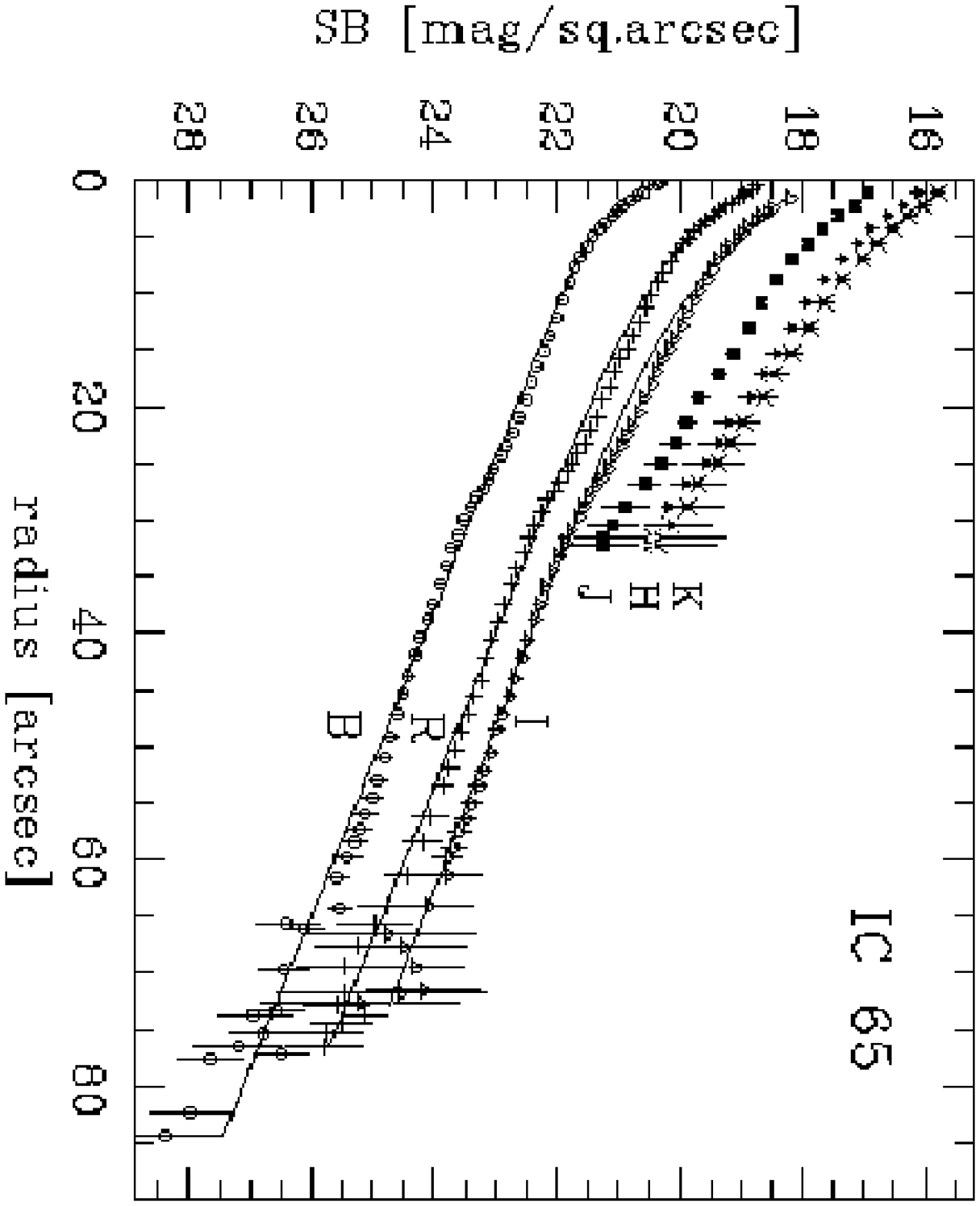}
\includegraphics{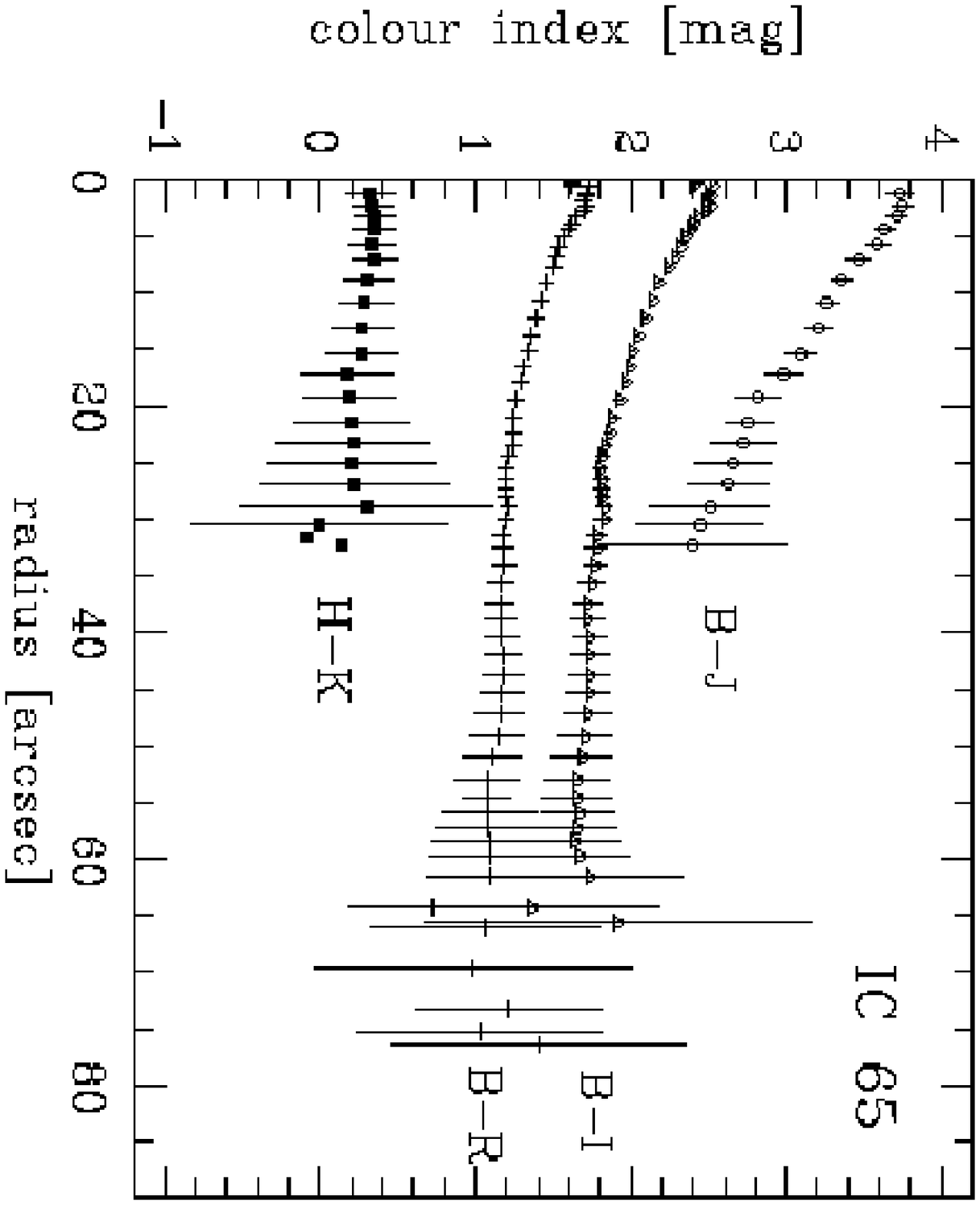}
\end{picture}
\caption{ IC 65 - parent galaxy of the group. {\it Upper left:} $B$-band image depicts an elliptical
central bulge and/or a bar, at the ends of which a long symmetric double-armed spiral pattern starts,
which includes many luminous knots and patches -- HII regions.
{\it Upper center:} $B-R$ color-index image shows that knotty spiral arms are blue (i.e. light shade),
with typical colors in the range of $0.7 - 0.95$ mag. The central elliptical bulge and the bar are red
(dark shade) with
$<B-R>^c\simeq$ 1.55, i.e. this component consists of old stars and dust.
{\it Upper right:} the 2MASS $JHK$ composite image appears smoother than the optical image; it is dominated
by an old stellar population and less affected by dust. $K$-band Laplacian contours (white)
delineate the nucleus and the bar as well as two round features beyond the end of the bar, which may be
intersections of an inner tilted ring with the plane of sky. 
The center of the fitted free ellipses is systematically drifting to the SE, both in optical and NIR bands
(decentering $\sim$ 8 \%, see text).
This is opposite to the global HI distribution, which shows an extension to the NW (see Fig.~1).
{\it Lower left:} a small bulge and/or a nucleus (within $\sim 10''$) and a bar component could be identified
both in optical and NIR SB profiles, followed by an extended nearly exponential disk, visible in optical 
bands.The B+D fit to the optical SB profiles is denoted by a continuous line.
{\it Lower right:} the optical and NIR color-index profiles show red bulge + bar area within $\sim 25''$,
followed by a nearly homogeneous disk, but still with some blueing towards the periphery, which is an effect of
blue spiral arms. Large gradients in the optical minus NIR colors probably show the occurence of dust
in the bar region.
}
\end{figure*}

\newpage
\begin{figure*}[hp]
\unitlength0.1cm
\begin{picture}(160,170)
\includegraphics{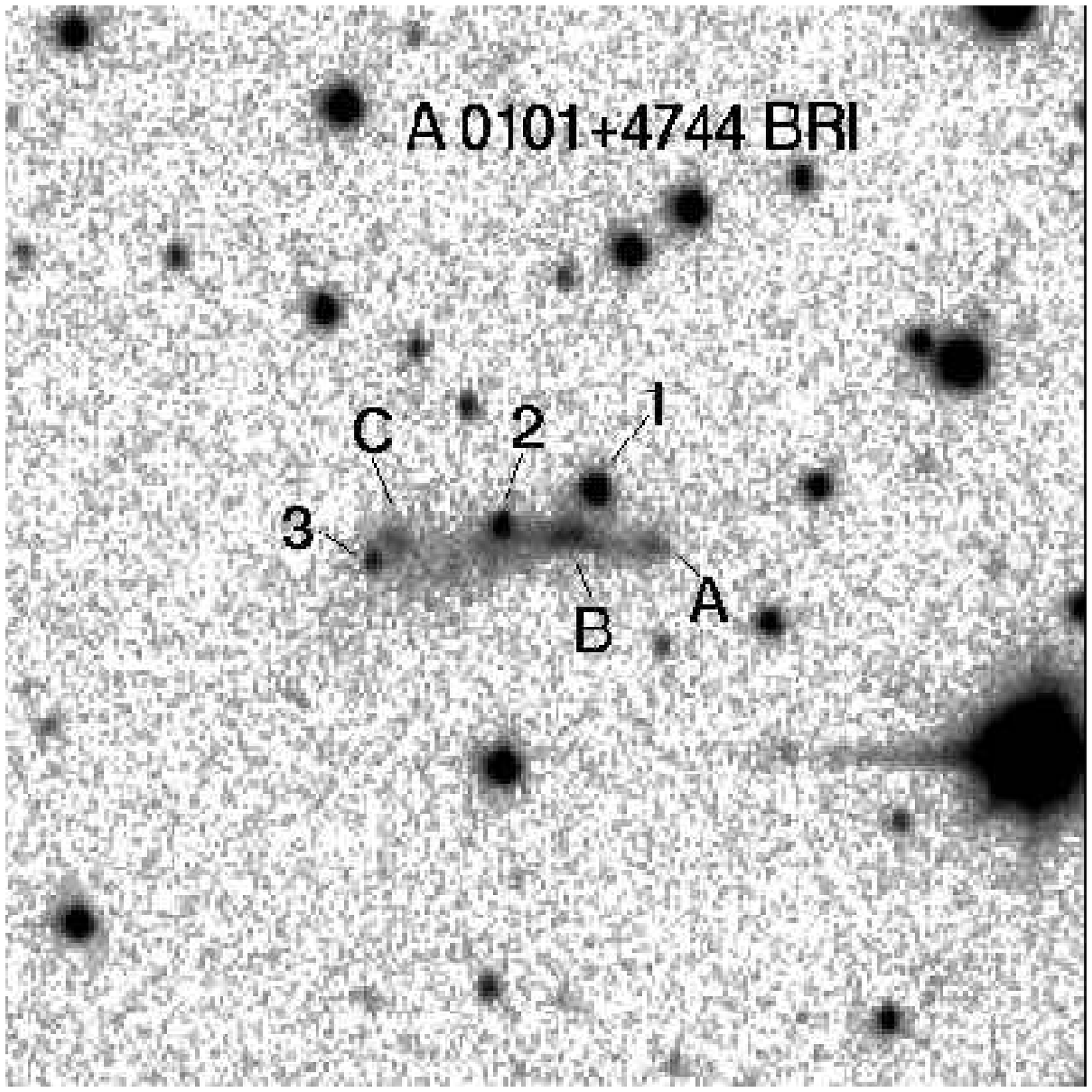}
\includegraphics{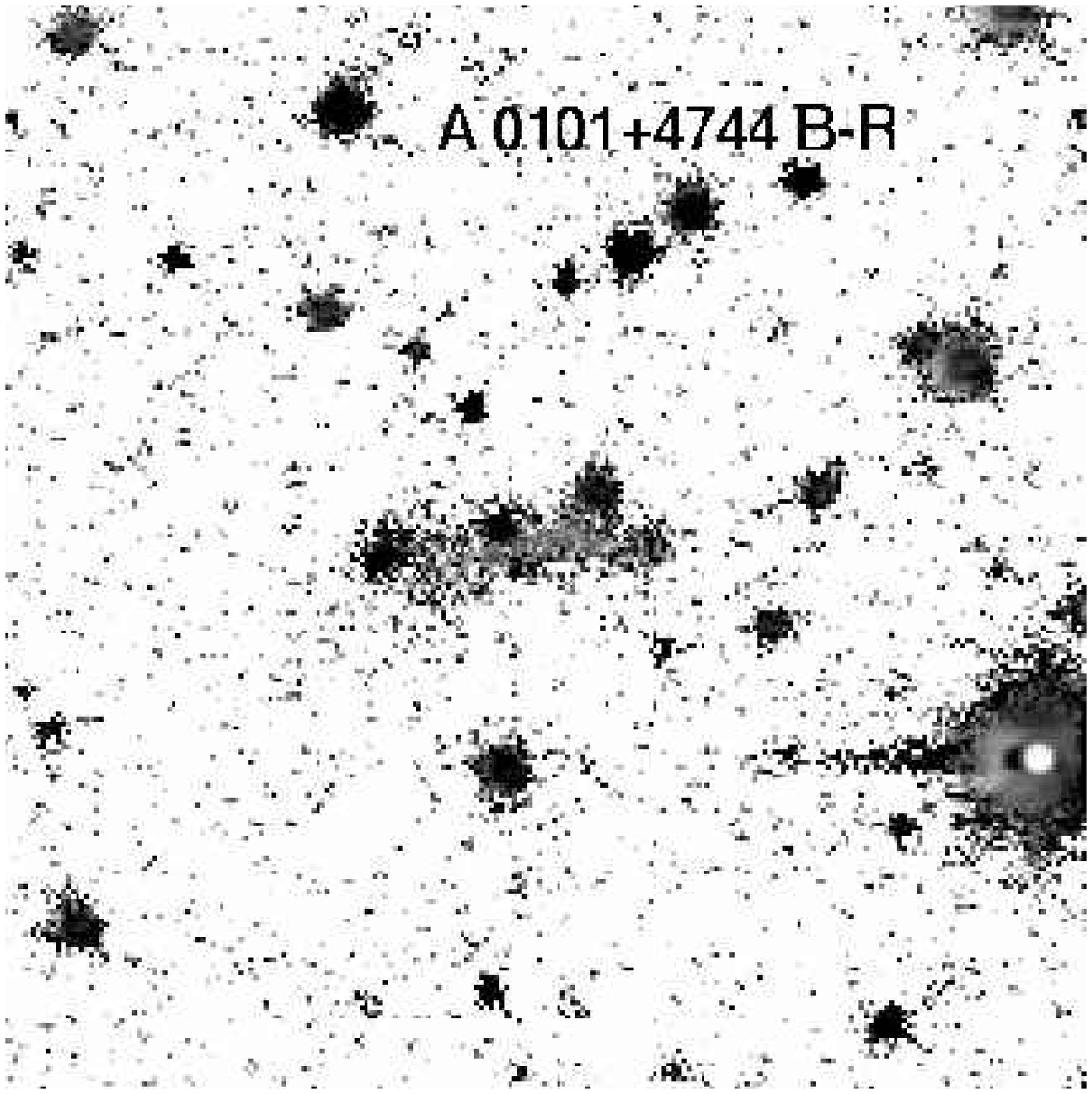}
\includegraphics{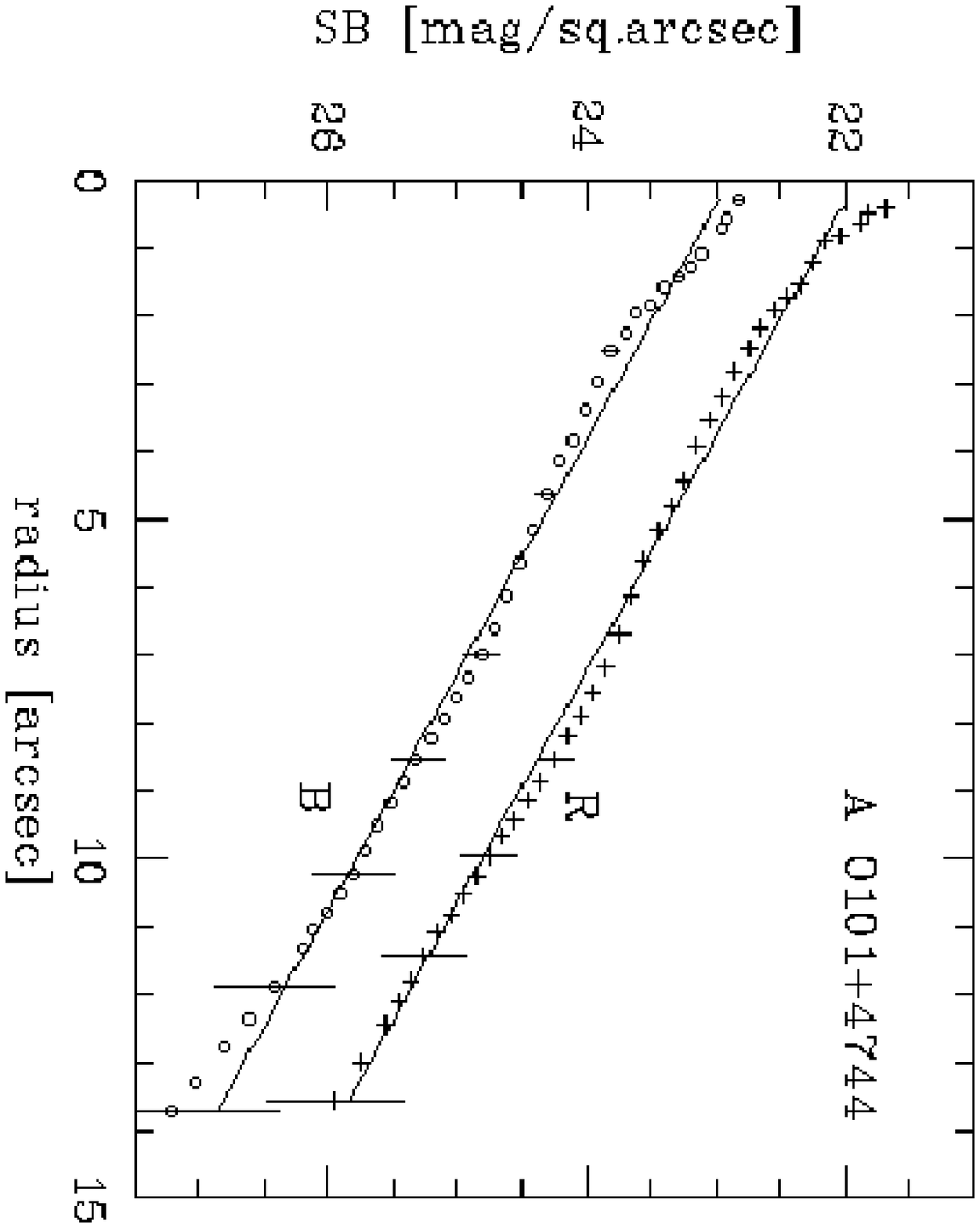}
\includegraphics{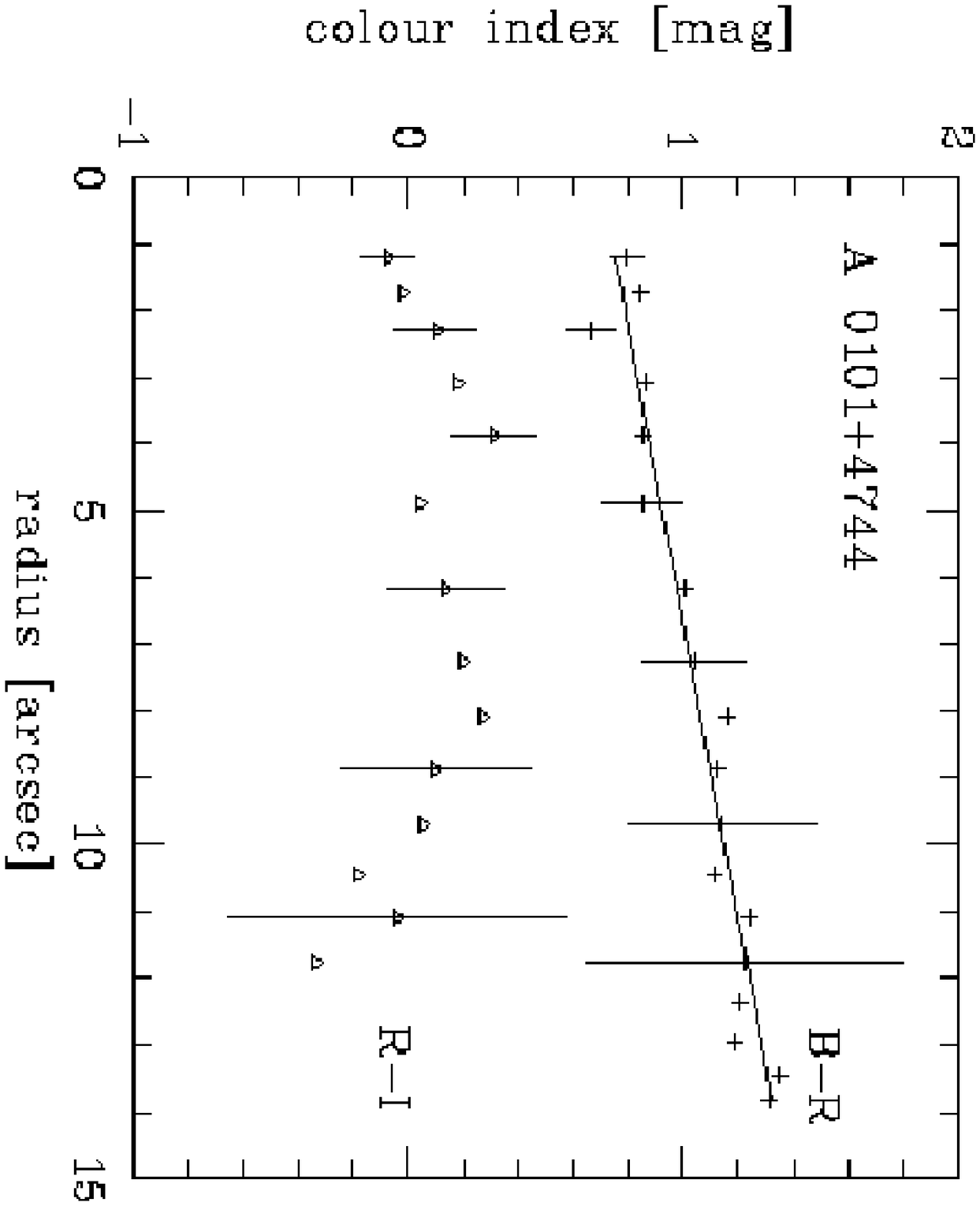}

\end{picture}
\caption{
{\it Upper left:} an anonymous irregular LSB galaxy A 0101+4744 of cometary appearance, 
with its brightest and bluest ($B-R = 0.65, R-I = 0.1$) star-forming (SF) knot {\bf B} 
located in the western part and a diffuse curved tail on the opposite side. 
Other diffuse knots {\bf A} and {\bf C} are redder ($<B-R>^c = 1.0 - 1.2$). 
We could distinguish galactic stars ({\bf 1, 2} and {\bf 3}) 
from SF knots by their stellar PSF 
and/or their typically redder colors.
{\it Upper right:} $B-R$ color-index image in the range 0.0 - 1.5 mag; the dark shade is red, the  
light shade is blue. Pixels with counts less than 2$\sigma$ of the background are flagged out.  
{\it Lower left:} light distribution of the underlying stellar 
component is nearly exponential. 
{\it Lower right:} $B-R$ color-index profile (centered on knot {\bf B}) shows 
significant reddening towards the periphery. 
The HI radial velocity, measured by van Moorsel (1983), confirms the  
membership of this particular dwarf galaxy ($M_B = -15.4, D_{25} \simeq 5$ kpc, exp. sc. length $h 
\simeq$ 0.7 kpc) in this group.  
} 
\end{figure*}

\begin{figure*}[hp]
\unitlength0.1cm
\begin{picture}(155,100)
\protect\label{fig5}
\includegraphics{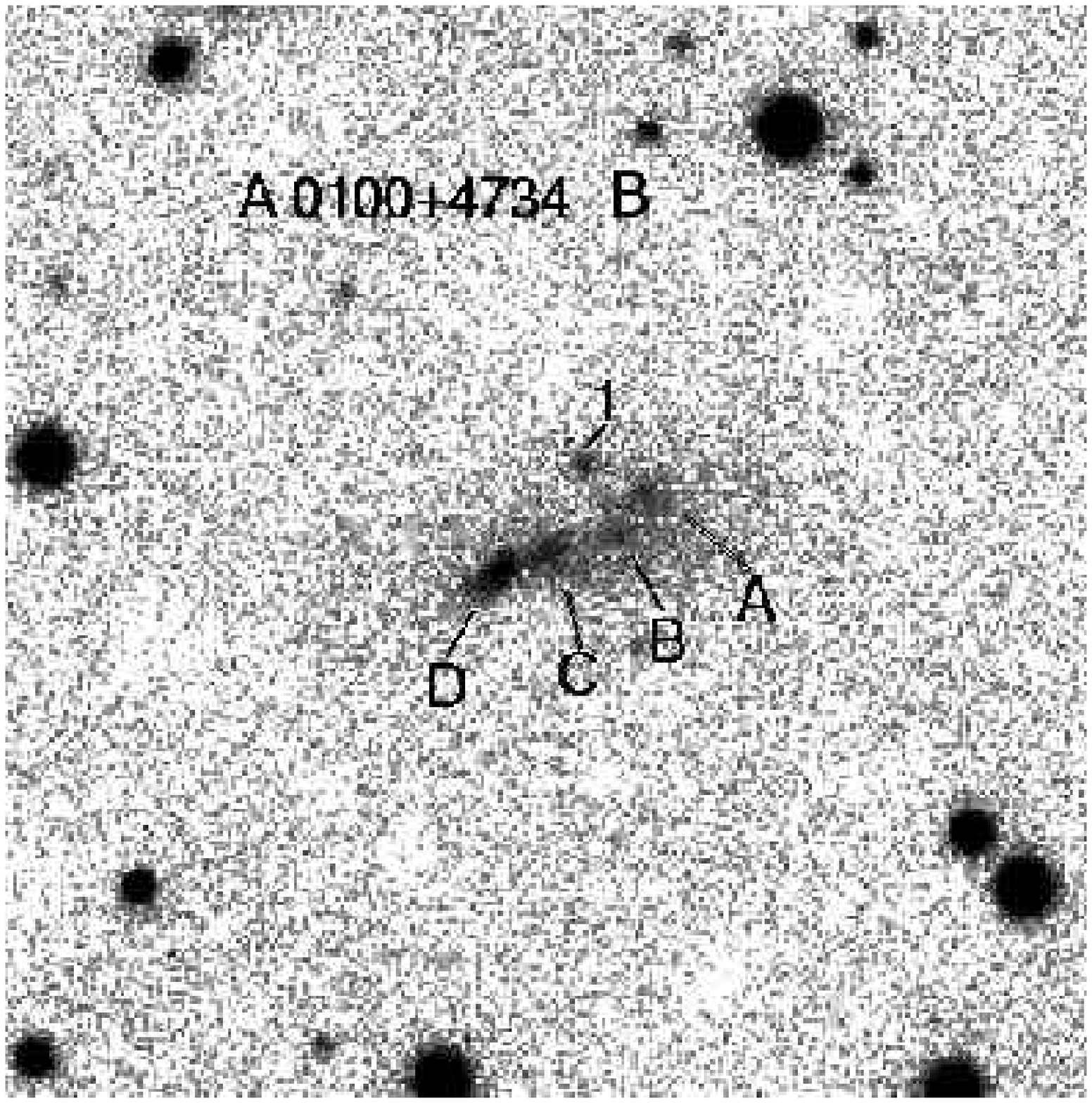}
\includegraphics{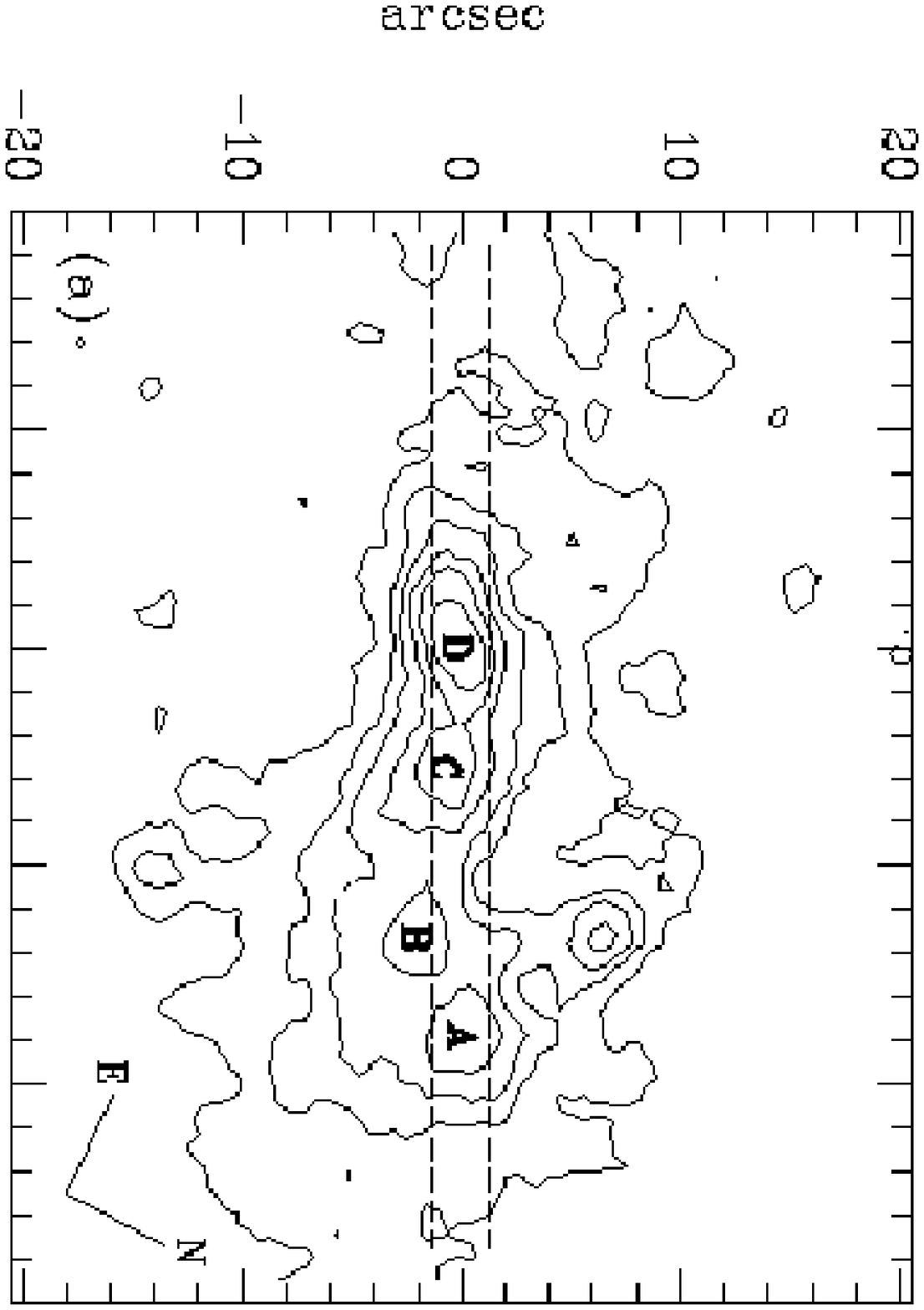}
\includegraphics{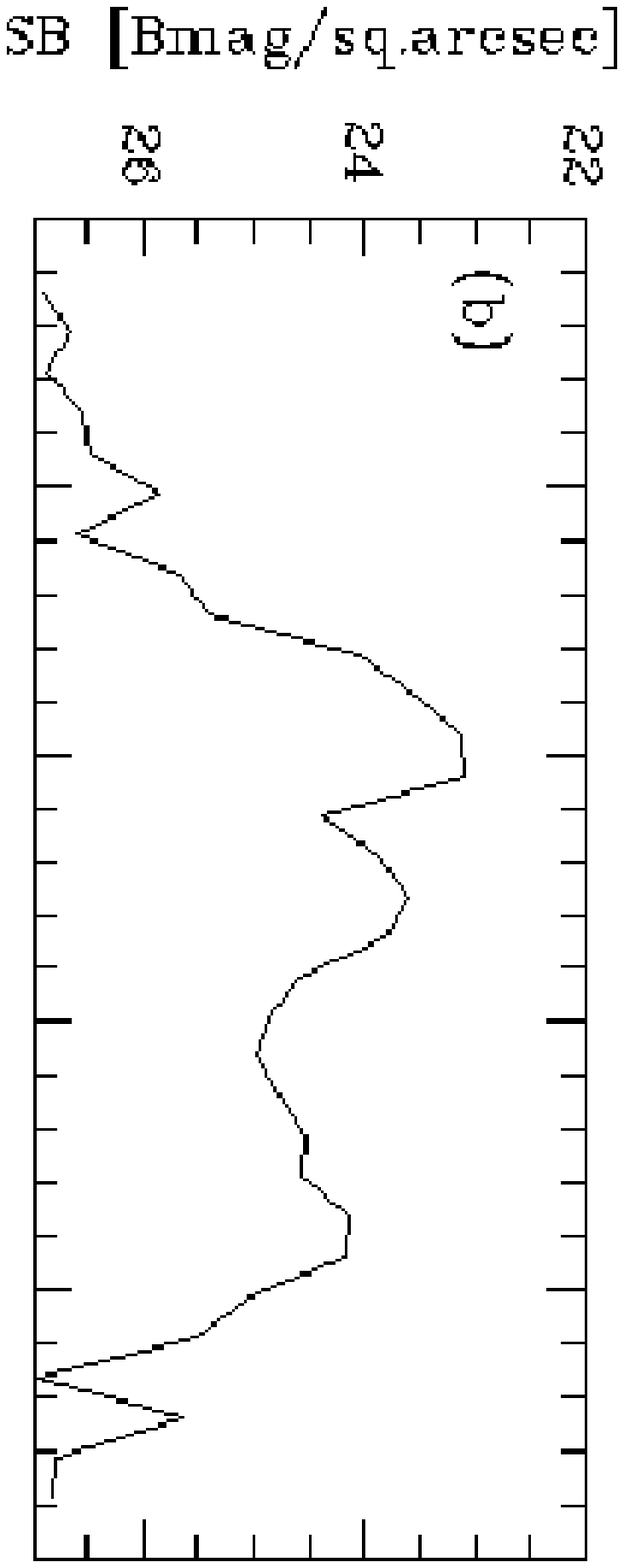}
\includegraphics{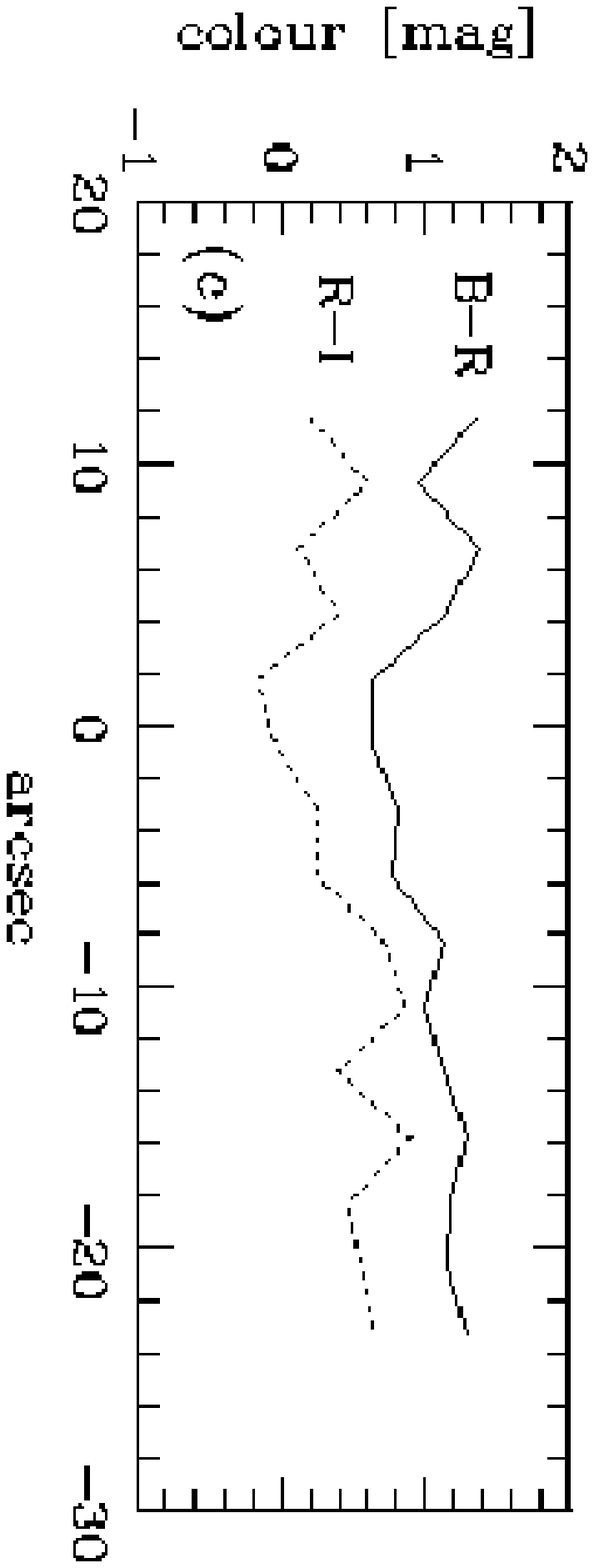}

\end{picture}
\caption{ {\it Left:} an anonymous irregular LSB galaxy A 0100+4734, located $\sim 10'$ to the SW of IC 65; 
 consists of a chain of four (blue) 
SF knots ({\bf A, B, C} and {\bf D}) with the brightest and the bluest knot {\bf D} located 
at the eastern periphery and, therefore, giving the galaxy a cometary appearance. Galactic star - 
{\bf 1}, projecting to the diffuse halo of the galaxy, is much redder than the SF knots.
{\it Right:} a $2.5''$ wide rectangular strip, extracted along the major axis intersects all major SF knots, 
which reveal a range of colors with fainter knots getting redder along the distance 
from the brightest/bluest knot {\bf D}. These color variations could be interpreted either as an 
age effect in the self-propagating SF mode or as an effect of inhomogenities in the dust distribution. 
The galaxy has been marginally detected in HI during the pilot observations, made by Huchtmeier (2002) 
in July 2002 with Effelberg's 100-m radio telescope (HPBW = 9.3$'$) in the redshift range of the group. 
However, this detection should be confirmed  
with better angular resolution.  }
\end{figure*}

\newpage 
\begin{figure*}[hp]
\unitlength0.1cm
\begin{picture}(155,100)
\includegraphics{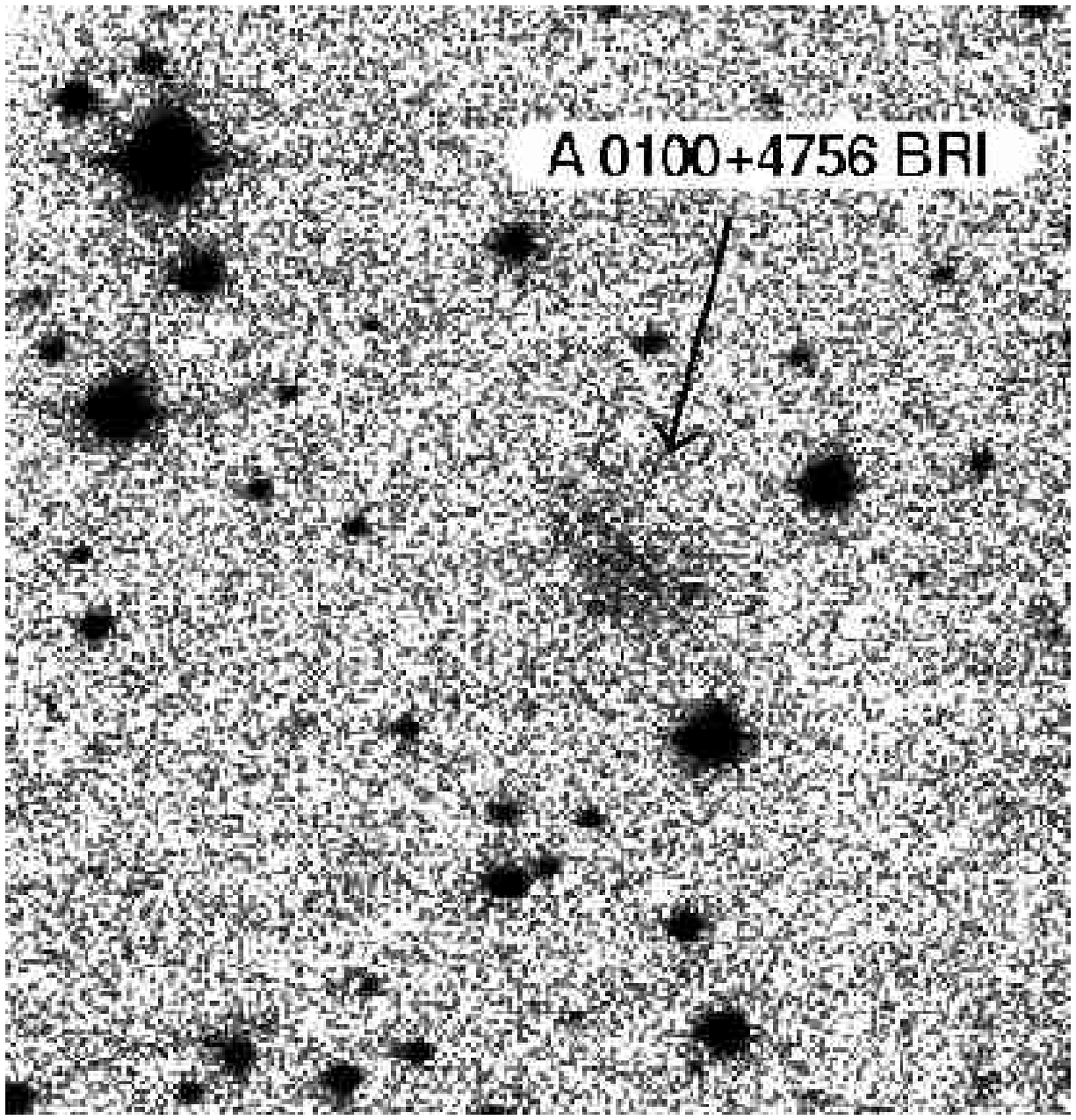}
\includegraphics{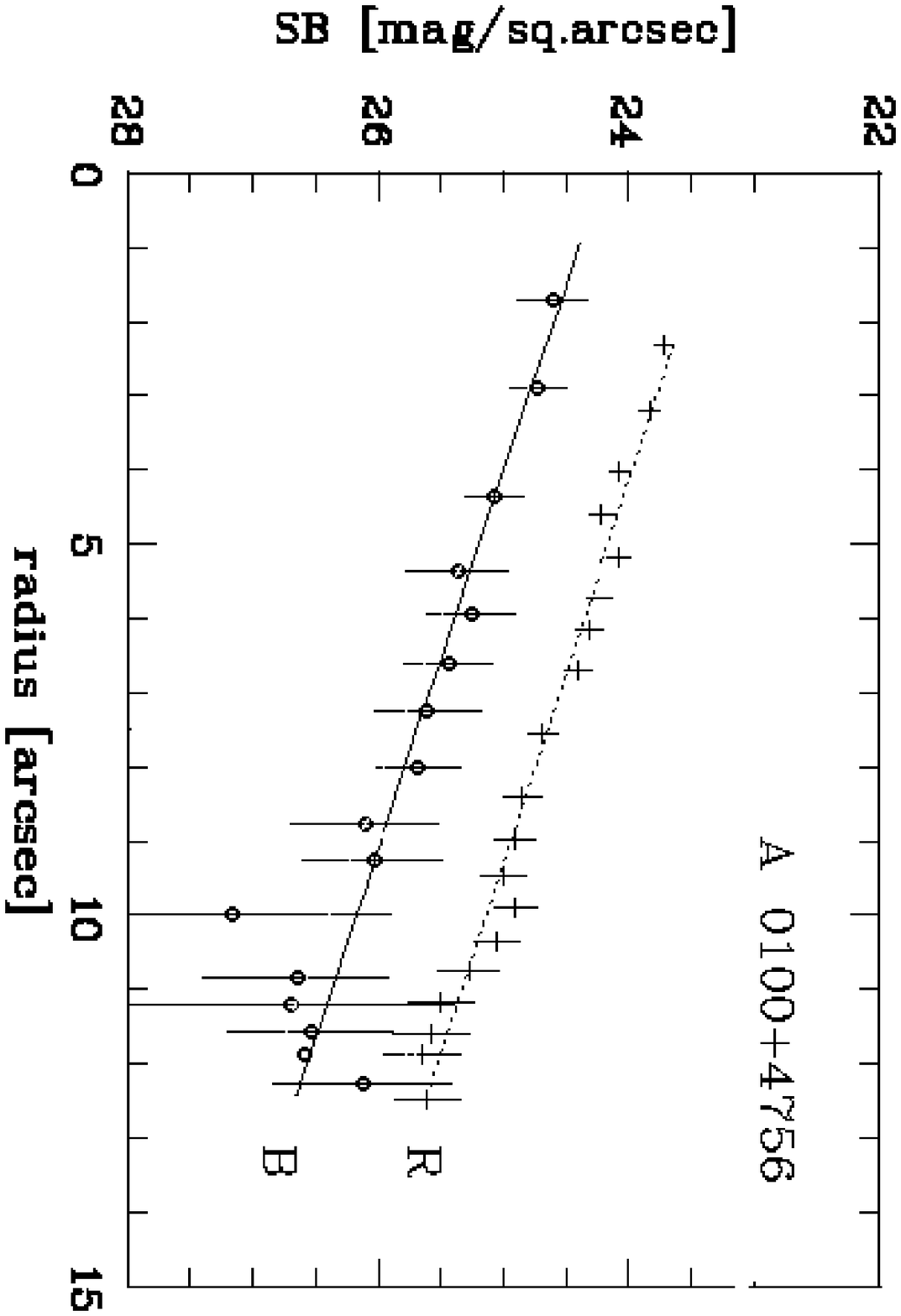}
\includegraphics{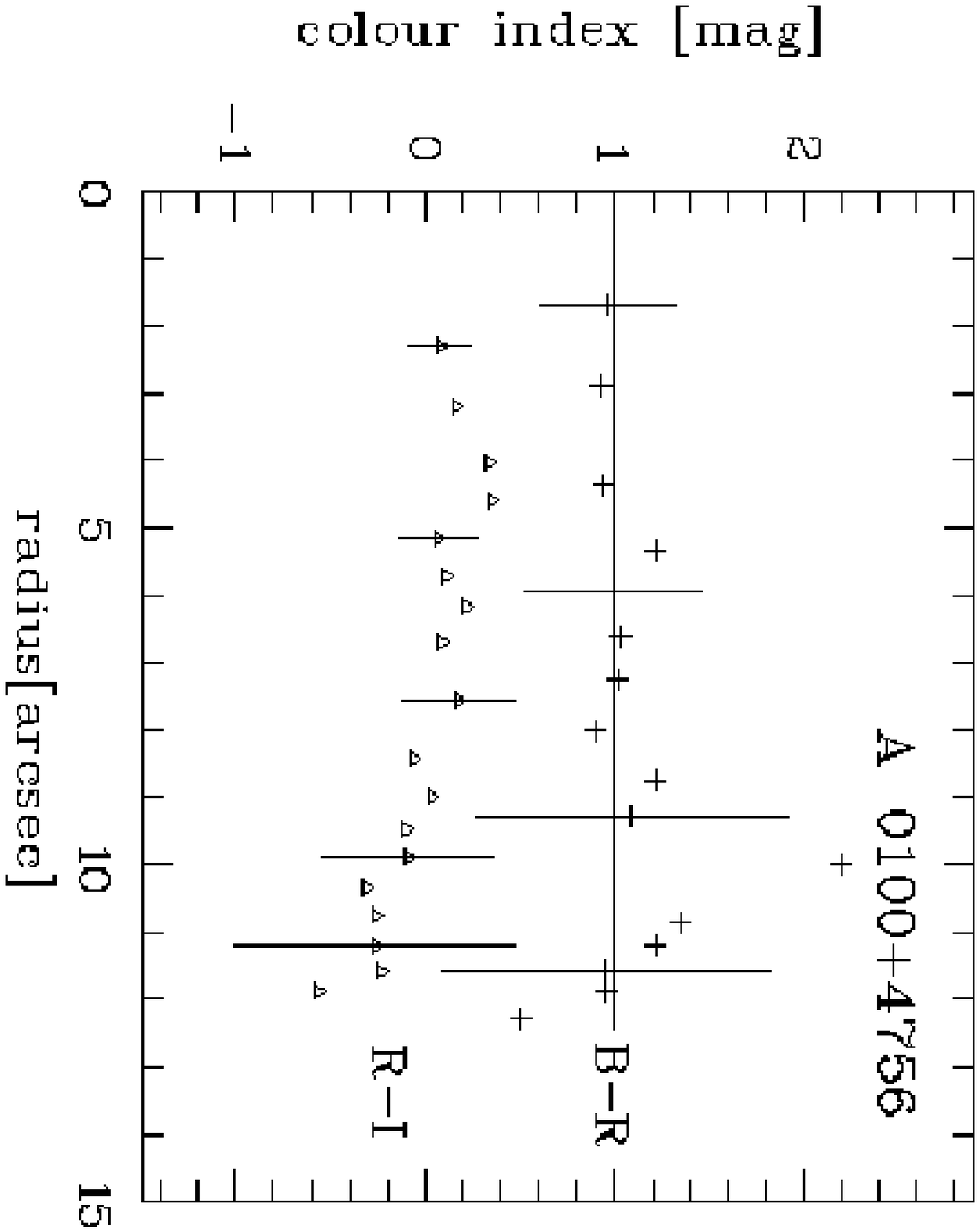}

\end{picture}
\caption{ A very faint LSB irregular object A 0100+4756 
showing a smooth light distribution ({\it left panel}) with a maximum 
SB $\simeq 24.0~B/\Box''$, exponential light profile ({\it upper right}) and with homogeneous blue 
($<B-R>^c \simeq 0.76 \pm 0.2, <R-I>^c \simeq 0.15 \pm 0.2$) color profile ({\it lower right}). 
The color characteristics 
are similar to those of A 0101+4744 -- the confirmed dwarf member of this group. The dwarfish 
morphology and location close to the UGC 622 ($\sim 5'$) let us consider this object for 
a possible member of this group. In this case, its main photometric parameters are: $M_B = -14.4, 
D \simeq 3.6$ kpc, $h \simeq$ 1.0 kpc.
Alternatively, this peculiar object could also be an isolated fragment of a faint Galactic 
reflection nebula.
}
\end{figure*}

\begin{figure*}[hp]
\unitlength0.1cm
\begin{picture}(155,100)
\includegraphics{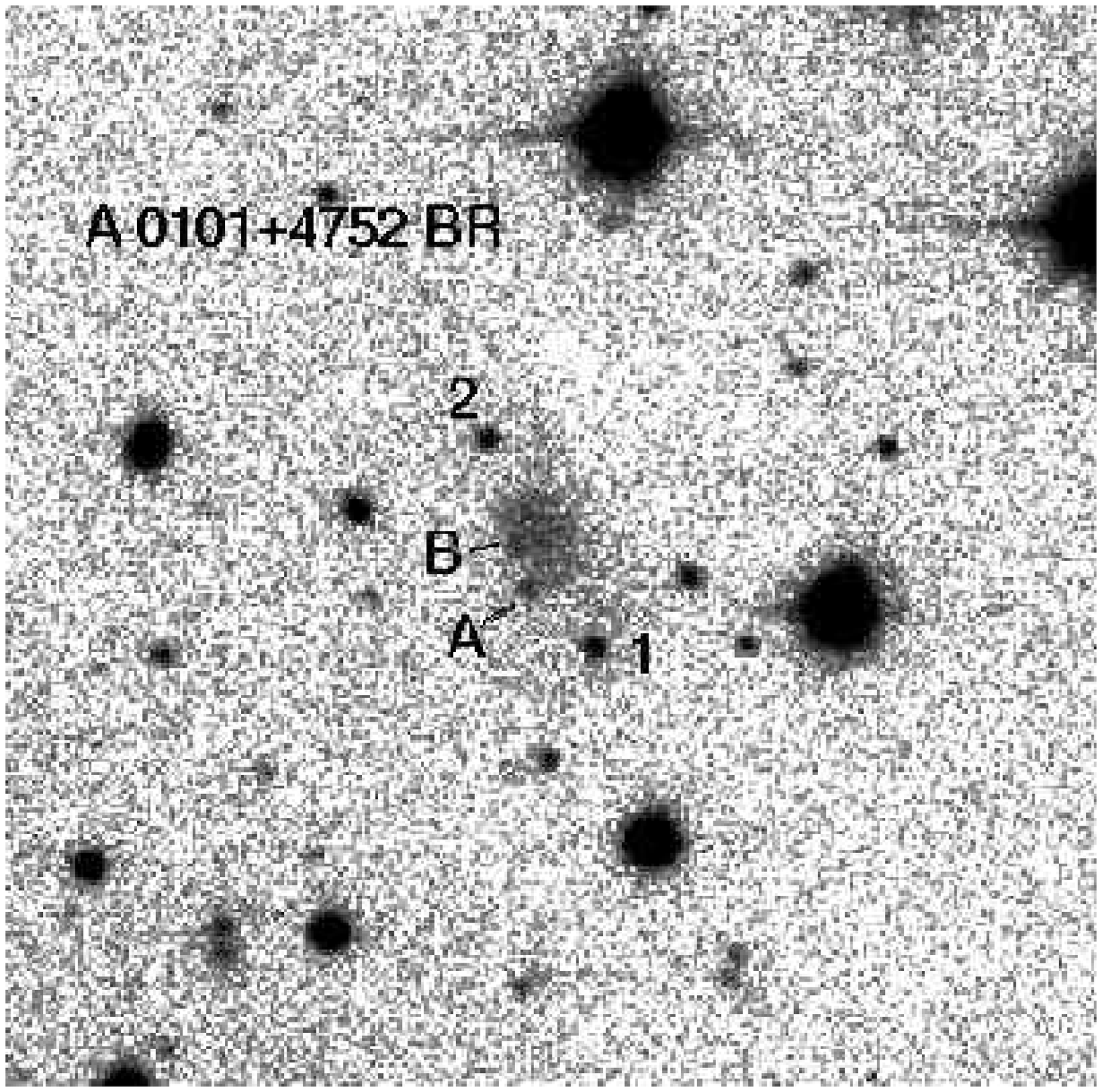}
\includegraphics{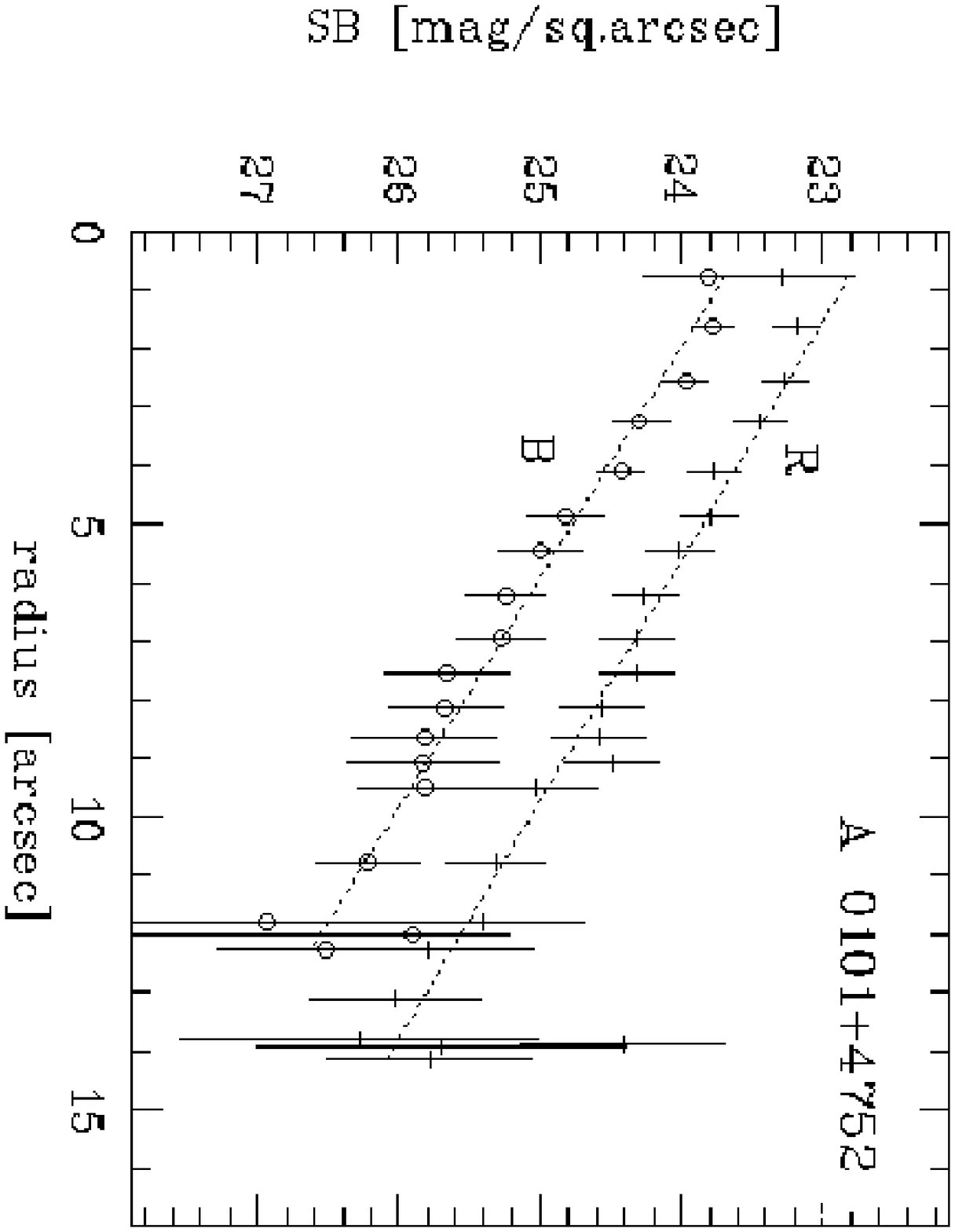}
\includegraphics{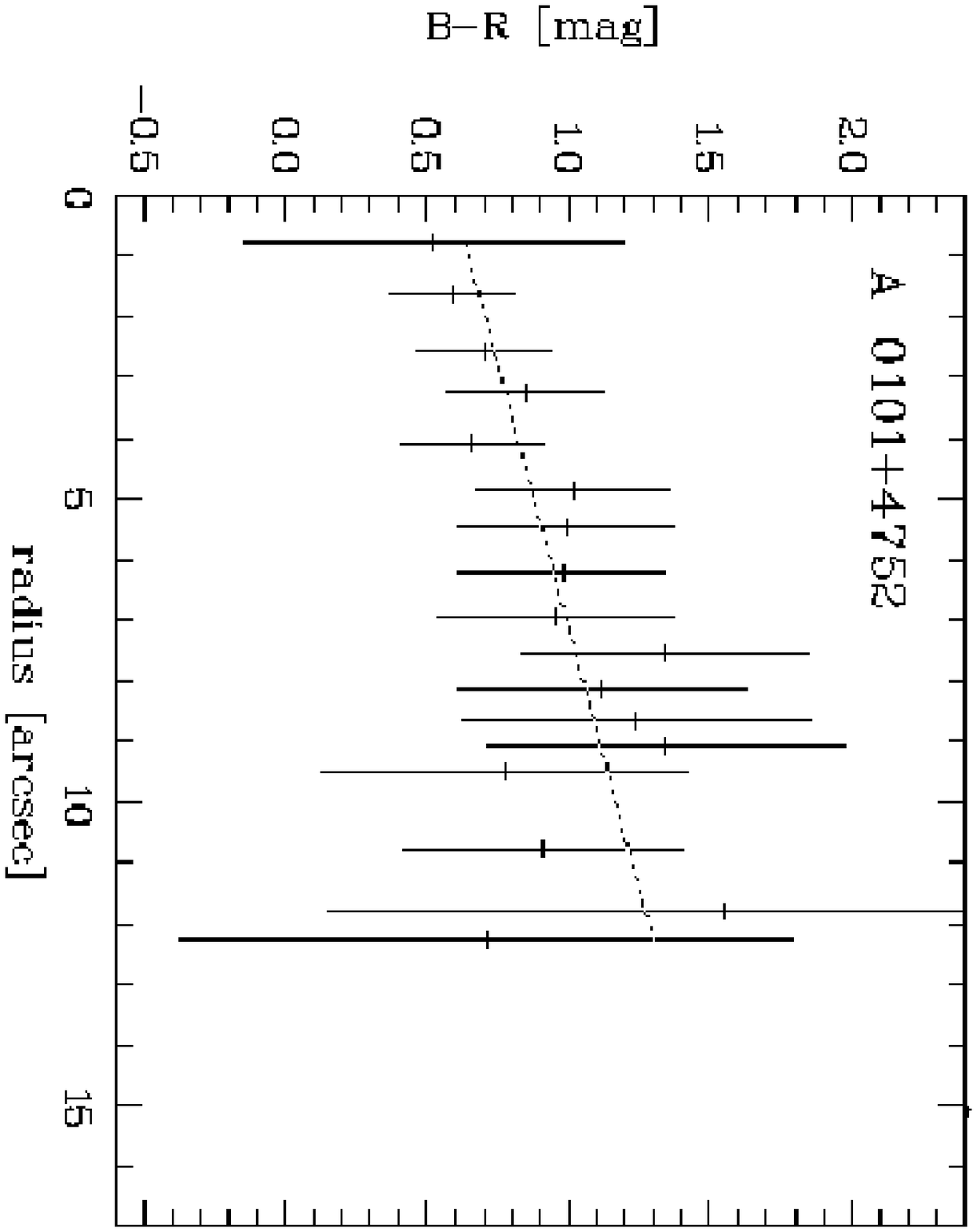}
\end{picture}
\caption{ An irregular LSB galaxy A 0101+4752 with two faint arc-like features, one streching 
to the North with the (red) stellar knot {\bf 2} at its end and another shorter arc ending at a resolved 
knot {\bf A}. Both arcs, marginally visible on the $BR$ composite CCD image ({\it left panel}), are confirmed by 
deeper (but low resolution) DSS 2 frames. Another resolved knot {\bf B} is extremely blue: $(B-R)^c 
\simeq$ 0.3. Light distribution of the underlying stellar component is nearly exponential 
({\it upper right}). 
The $B-R$ color-index profile ({\it lower right}), starting from the bluest knot {\bf B} shows clear reddening towards 
the periphery. Morphological features, as the absence of the regular spiral structure and 
the occurrence of several 
blue SF (?) knots let us consider this object as a relatively nearby dwarf galaxy. When associated with 
the IC 65 group, its photometric characteristics ( $M_B = -14.9, D \simeq 4.5$ kpc, 
$h \simeq 0.8$ kpc, $<B-R>^c \simeq 0.7, <R-I>^c \simeq 0.2$) are rather typical for an irregular 
dwarf galaxy with slow star formation. 
}
\end{figure*}

\end{document}